\documentclass[aps,prl,twocolumn,superscriptaddress,showpacs]{revtex4}
\usepackage{graphicx}
\begin{document}

\title{Coulomb drag by small momentum transfer between quantum wires}

\author{M. Pustilnik}
\affiliation{Theoretical Physics Institute, University of Minnesota, 
Minneapolis, MN 55455}
\affiliation{School of Physics, Georgia Institute of Technology, 
Atlanta, GA 30332}
\author{E.G. Mishchenko}
\affiliation{Bell Laboratories, Lucent Technologies, 
Murray Hill, NJ 07974} 
\affiliation{Department of Physics, University of Colorado, 
Boulder, CO 80390}
\author{L.I. Glazman}
\affiliation{Theoretical Physics Institute, University of Minnesota, 
Minneapolis, MN 55455}
\author{A.V. Andreev}
\affiliation{Bell Laboratories, Lucent Technologies, Murray Hill, NJ 07974}   
\affiliation{Department of Physics, University of Colorado, Boulder, CO 80390}

\begin{abstract}
We demonstrate that in a wide range of temperatures Coulomb drag
between two weakly coupled quantum wires is dominated by processes
with a small interwire momentum transfer. Such processes, not
accounted for in the conventional Luttinger liquid theory, cause drag
only because the electron dispersion relation is not linear. The
corresponding contribution to the drag resistance scales with
temperature as $T^2$ if the wires are identical, and as $T^5$ if the
wires are different.
\end{abstract}
\pacs{
71.10.Pm,   %Fermions in reduced dimensions
72.15.Nj,   %Collective modes (e.g., in one-dimensional conductors)   
73.23.Ad,    %Ballistic transport   
73.63.Nm    %Quantum wires (Electronic transport...)
\vspace{-3mm}
}
\maketitle

\begin{figure}[h]
\includegraphics[width=0.5\columnwidth]{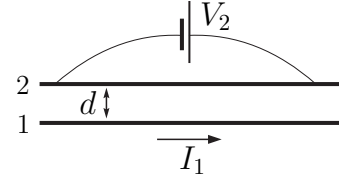}
\vspace{-1mm}
\caption{Coulomb drag between quantum wires. A dc current $I_1$ 
flows through the active wire (1). A voltage bias $V_2$ is applied to the 
passive wire (2) in such a way that $I_2 = 0$.
\label{drag_fig}
}
\vspace{-2mm}
\end{figure}

Electrons moving in a conductor generate a fluctuating electric field
around it. This field gives rise to an unusual transport phenomenon,
Coulomb drag between two closely situated conductors~\cite{Rojo}. The
structure of the fluctuating field is determined by electron
correlations within the conductors.  Correlations are stronger in
conductors of lower dimensionality. Tomonaga-Luttinger model captures
some aspects of the correlations in the case of one-dimensional
conductors (quantum wires). Within this model, the Coulomb drag was
studied in~\cite{NA,KS}. In a typical setup~\cite{exp1,exp2} a dc
current flows through the active wire $1$, while the bias applied to
the passive wire sets $I_2=0$, see Fig.~\ref{drag_fig}. The
\textit{drag resistivity} (drag resistance per unit length of the
interacting region) is then defined as
\begin{equation}
r = - \lim_{I_1\to 0} \frac{e^2}{2\pi\hbar}\frac {1}{L} \frac{dV_2}{dI_1} .
\label{def} 
\end{equation}

The only source of drag in the Luttinger liquid is interwire
backscattering, associated with a large momentum transfer between the
wires. The model predicts a distinctive temperature dependence of the
corresponding contribution $r_{2k_F}$ to the drag
resistivity~(\ref{def}). In the case of identical wires $r_{2k_F}
\propto l_{2k_F}^{-1}e^{\Delta/T}$ at the lowest
temperatures~\cite{NA,KS}. Here $l_{2k_F}$ is the scattering length
characterizing the interwire backscattering. At temperatures $T$ above
the gap $\Delta$, this exponential dependence is replaced by a
power-law, $r_{2k_F} \sim l_{2k_F}^{-1} (T/\epsilon_F)^{1-\gamma}$,
where $\epsilon_F$ is the Fermi energy.  The exponential temperature
dependence of $r_{2k_F}$ indicates a formation of a zig-zag charge
order due to the $2k_F$-component of the interwire
interaction~\cite{NA,KS}.  To the contrary, the exponent $\gamma >0$
in the power-law portion of the function $r_{2k_F} (T)$ is determined
by the interactions within the wires; $\gamma=0$ in the absence of
interactions~\cite{HF}. This renormalization of $r_{2k_F}$
is similar in origin to the suppression of the conductance of a Luttinger 
liquid with an impurity~\cite{Kane_Fisher}: in both cases repulsive interactions
enhance the backscattering probability when temperature is lowered.

However, \textit{forward} scattering between the wires also induces
drag. To see this, one has to go beyond the Tomonaga-Luttinger model and
account for the nonlinearity of the electronic dispersion relation.
If the electron velocity depends on momentum, then even small
(compared to $2k_F$) momentum transfer results in drag.

%%%%%%%%%%%%%%%%%%%%%%%%%
%%%%%%%%%%%%%%%%%%%%%%%%%

The small ($r_0$) and large ($r_{2k_F}$) momentum transfer
contributions to the drag are inversely proportional to the
corresponding scattering lengths $l_0$ and $l_{2k_F}$,
respectively. Their ratio $l_{0}/l_{2k_F}\propto e^{-4k_Fd}$ depends
strongly on the distance $d$ between the wires. If the drag effect is
employed to study the correlations within a wire, rather than the zig-zag
order induced by interwire interaction, then $d$ must be large: 
$k_Fd>1$. In this case the gap $\Delta\sim\epsilon_F(k_F
l_{2k_F})^{-1/\gamma}$ becomes narrow, and the role of forward scattering
increases.

In this paper we demonstrate that the drag resistivity between weakly
coupled wires is dominated by the forward scattering in a wide 
temperature range. Even for identical wires, which is the most favorable 
for backscattering case, $r_0\propto T^2$ wins over $r_{2k_F}$ at all 
$T$ above $ T^*\sim \epsilon_F (l_0/l_{2k_F})^{1/(1+\gamma)}$.  
For different wires, $r_{2k_F}$ is exponentially small at $T\lesssim u\delta n$, 
whereas $r_0$ has a power-law low-temperature asymptotics; here $\delta n$ 
is the mismatch of the electron densities between the wires and $u$ is the 
characteristic plasma velocity; hereafter we set $\hbar =1$.

The Hamiltonian of the system, $H = H_1 +H_2 + H_{12}$,
is given by the sum of the Hamiltonians of the two isolated wires $i=1,2$,
\begin{eqnarray}
&& H_i = \int dx \,\psi_i^\dagger(x) \frac{\hat p^2}{2m} \psi_i (x) +
H_{int},
\quad \hat p = -id/dx, 
\nonumber\\
&& H_{int} = \int dxdy \, \rho_i (x) U_{ii} (x-y) \rho_i (y),
\label{H_i}
\end{eqnarray} 
and of the Hamiltonian of the interwire interaction, 
\begin{equation}
H_{12} = \int dx_1 dx_2 \, \rho_1 (x_1) U_{12}(x_1- x_2) \rho_2 (x_2);
\label{H_12}
\end{equation}
here $\rho_i(x) = \psi_i^\dagger(x) \psi_i(x)$.  

We will assume that the interactions are screened by the nearby
metallic gates.  The screening length $d_s$ is set by the distance to
the gates and is typically~\cite{exp1,exp2} of the order of the
separation $d$ between the wires, $d_s\sim d$. The short-distance
cutoff $d_{ij}$ of the Coulomb potential is provided by the wire width
$d_0$ for $i=j$ or by the interwire distance $d$ for $i\neq j$. The 
Fourier transforms $U_{ij}(k)= \int dx e^{ikx}U_{ij}(x)$ are rapidly 
decreasing functions of $k$ with $U_{ij}(k)\propto e^{-|k| d_s}$ for 
$|k| \gg 1/d_s$, and $U_{ij}(k) \approx {\rm const}$ for $|k|\ll 1/d_{ij}$.  
Note that $d_{ij} \sim d_s \sim d$ for interwire interaction $U_{12}(k)$.  
Thus, its $k$-dependence is characterized by a single scale $k_0\sim 1/d$.

Because of the interaction $H_{12}$, electrons in the wire $2$
experience a force~\cite{NA} whose density is given by
\begin{equation}
{\cal F}_2  
= \int dx \left(dU_{12}(x)/dx\right)\langle \rho_1 (x)\rho_2 (0)
\rangle .
\label{force}
\end{equation}
Since there is no current in the wire 2, this force must be balanced
by an external electric field, $e n_2 {\cal E}_2 + {\cal F}_2 = 0$,
where ${\cal E}_2 = V_2/ L$ and $n_i = \langle \rho_i\rangle$ is the
concentration of electrons in the wire $i$.  At $T \gg \Delta$ (see
the discussion above) the correlation function in the r.h.s. of
Eq.~(\ref{force}) can be evaluated in the first order in $U_{12}$,
\begin{equation}
\frac{V_2}{L} = \frac{1}{en_2} \int
\frac{dk d\omega }{(2\pi)^2}~
k U_{12}^2(k) \widetilde S_1(k,\omega) \widetilde S_2 (-k,-\omega),
\label{voltage}
\end{equation}
where $\widetilde S_i(k,\omega)$ are the dynamic structure factors,
\[  
\widetilde S_i (k,\omega)  = \int dx dt ~e^{i\omega t - ikx}
\left\langle \rho_i (x,t)  \rho_i (0,0)\right\rangle, 
\]
calculated in the presence of a finite current $I_1$ in the wire $1$.
The structure factor $\widetilde S_2(k,\omega)$ in the wire 2
coincides with its equilibrium value, $S_2(k,\omega)$.  The electronic
subsystem in the wire $1$ is in equilibrium in the reference frame
moving with the drift velocity $v_d= I_1/e n_1$ in the direction of
the current.  Therefore the structure factor $\widetilde S_1$ is
obtained from the equilibrium value $ S_1$ using the Galilean
transformation: $\widetilde S_1(k,\omega) = S_1(k,\omega-qv_d)$.
Equations~(\ref{def}) and (\ref{voltage}) then yield
\begin{equation}
r = \int dkd\omega~\frac{k^2 U^2_{12}(k)}{8\pi^3 n_1 n_2} 
\frac{\partial S_1(k,\omega)}{\partial\omega} S_2(-k,-\omega) .
\label{drag0}
\end{equation}
Now we use the fluctuation-dissipation theorem,
\[
S_i (k,\omega) = \frac{2A_i(k,\omega)}{1-e^{-\omega/T}}
\]
to further simplify Eq.~(\ref{drag0}),
\begin{equation}
r  = \int_0^\infty dk \int_0^\infty d\omega~
\frac{k^2 U_{12}^2(k)}{4\pi^3 n_1n_2 T} 
\frac{A_1(k,\omega) A_2(k,\omega)}
{\sinh^2(\omega/2T)}.
\label{drag}
\end{equation}
Here $A_i$ is the imaginary part of the retarded density-density
correlation function; $A_i(k,\omega) = A_i(-k,\omega) = - A_i(k,-
\omega)$. Equation~(\ref{drag}) was derived by different means
in~\cite{ZM}; similar expressions have been also obtained for
noninteracting systems with disorder~\cite{nonint}. Here we
demonstrated the validity of Eq.~(\ref{drag}) for clean interacting
systems.

We start with the evaluation of the drag resistivity for
noninteracting electrons ($U_{ii}=0$). Concentrating on the small
momentum transfer contribution to $r$, we consider the limit
$l_0/l_{2k_F}\to 0$, thus setting $r_{2k_F}=0$. In this case the main
contribution to the integral over $k$ in Eq.~(\ref{drag}) comes from
small momenta $k\ll k_F$ and small energies $\omega\ll \epsilon_F$. At
these values of $k$ and $\omega$, functions $A_i(k,\omega)$ are
sharply peaked at $\omega_i = v_i k$, where $v_i = \pi n_i/m$ are the
Fermi velocities in the two wires. For a given $k<2k_F$ the widths of
the peaks can be estimated as
\begin{equation}
\delta\omega(k,T) 
=  {\rm max}\left\{ k^2/m, kT/k_F\right\} .
\label{delta_free} 
\end{equation} 
Equation (\ref{delta_free}) and the exact f-sum rule, 
\begin{equation}
\int_0^\infty d\omega~\omega A_i(k,\omega) = \frac{\pi n_i}{m}
\frac{k^2}{2} ,
\label{sum_rule}
\end{equation}
allow us to estimate the peak heights:
$A_i \sim k/2\delta\omega$.
If the difference between the Fermi velocities is small,  
\[
\delta v = |v_1-v_2| \ll v_F=\pi n/m,
\quad
n=(n_1+n_2)/2,
\]
then Eq.~(\ref{drag}) reduces to
\begin{eqnarray}
&& r  = \frac{1}{4\pi^3 n^2 T} 
\int_0^\infty dk ~
\frac{k^2 U_{12}^2(k)}
{\sinh^2(v_F k/2T)}\,
\alpha(k,T) , ~~~~
\label{drag_1}
\\
&&\alpha(k,T) = \int_0^\infty d\omega A_1(k,\omega)A_2(k,\omega) .
\label{alpha}
\end{eqnarray} 
The function $\alpha(k,T)$ depends on $\delta v$. If the
wires are identical ($\delta v = 0$), then Eq.~(\ref{alpha}) and the
above estimates for $A_i$ yield
\begin{equation}
\alpha(k,T) \approx \frac{k^2}{4\delta\omega(k,T)} .
\label{alpha_1}
\end{equation}
There are two competing scales in the integrand of Eq.~(\ref{drag_1}).
The first scale, $k_0\sim 1/d \ll k_F$, characterizes the $k$-dependence 
of the interwire interaction $U_{12}(k)$. The typical wave vector of thermally 
excited electron-hole pairs, $T/v_F$, defines the second scale. The two scales
coincide at $T=T_0 = v_F k_0$. At $T\ll T_0$ one can replace
$U_{12}(k)$ by $U_{12}(0)$ in Eq.~(\ref{drag_1}).  Furthemore, we 
use $\alpha$ in the form of Eq.~(\ref{alpha_1}) at $T=0$, which results in
\begin{equation}
r = \frac{c_1}{l_0}\left(\frac{T}{\epsilon_F}\right)^2,
\quad \frac{1}{l_0} = \left[\frac{U_{12}(0)}{2\pi v_F}\right]^2  n 
\label{drag_peak_free}
\end{equation} 
with $c_1 = \pi^4/12$.  Use of exact form of $A_i(k,\omega)$ in
Eq.~(\ref{alpha}) changes only the numerical coefficient, $c_1 =
\pi^2/4$.

The increase of temperature $T$ above $T_0$ results in a saturation of
the drag resistivity. Indeed, at $T_0 \ll T\ll \epsilon_F$ one can
expand $\sinh(v_F k/2T)$ in Eq.~(\ref{drag_1}) and use $\delta\omega =
kT/k_F$ for the peak width in (\ref{alpha_1}). This yields
\begin{equation}
r\sim \frac{1}{l_0} \int_0^\infty \frac {kdk}{n^2} 
\frac{U_{12}^2(k)}{U_{12}^2(0)}
\sim \frac{1}{l_0}\left(\frac{T_0}{\epsilon_F}\right)^2 .
\label{high_T}
\end{equation} 
Further increase of $T$ leads to the decay of the drag,
\begin{equation}
r \propto l_0^{-1} (T_0/\epsilon_F)^2(T/\epsilon_F)^{-3/2},\quad
T\gg\epsilon_F,
\label{hight}
\end{equation}
similar to the two-dimensional case~\cite{FH}. 

We now consider wires with slightly different Fermi velocities $\delta
v>0$.  In this case the peaks of $A_i(k,\omega)$ are separated in
$\omega$ by $k\delta v $. We define a new temperature scale $T_1 =
k_F\delta v$ by equating the separation to the peak width
(\ref{delta_free}). We assume this scale is small, $T_1 \ll T_0$.  The
difference between velocities does not affect the drag at $T\gg T_1$.
However, at $T\ll T_1$ the drag resistivity is suppressed
exponentially. To obtain the leading asymptotics of $r(T)$ it is
sufficient to use the $T=0$ limit~\cite{free} of $A_i(k,\omega)$ in
Eq.~(\ref{alpha}),
$\alpha(k,0) = 
(m/4k)\left(k-m\delta v\right)\theta(k-m\delta v)$.
Equation~(\ref{drag_1}) then results in
\begin{equation}
r = \frac{\pi^2/4}{l_0} \left(\frac{T_1}{\epsilon_F}\right)^2 
\frac{T}{T_1} e^{-T_1/T},
\quad
\quad T\ll T_1.
\label{off_peak_free} 
\end{equation} 

The activational temperature dependence Eq.~(\ref{off_peak_free})
holds for all $T\ll T_1$, because for noninteracting electrons at
$T=0$ the product $A_1(k,\omega) A_2(k,\omega)$ is exactly
zero~\cite{free} at $k<m\delta v$. If electrons interact, some overlap
of $A_1$ and $A_2$ exists even at small $k\ll m\delta v$. This yields
a further contribution to $r$, that has a power-law temperature
dependence. We will evaluate this contribution for weak intrawire
interaction.

It is convenient to write $A(k,\omega)$ (we suppress the index $i$ in
the following) in the form $A(k,\omega) = \left[S(k,\omega) -
  S(-k,-\omega)\right]/2$ and use the Lehmann representation for the
dynamic structure factor:
\begin{equation}
S(k,\omega) = \frac{2\pi }{L}
\sum_n \left|\left\langle n\right| \rho_k\left|gs\right\rangle\right|^2 
\delta(\omega - E_n + E_{gs}). 
\label{Lehmann}
\end{equation}
Here $L$ is the system size, $\left|gs\right\rangle$ is the ground
state, and $\rho_k = \sum_p \psi_{p+k}^\dagger \psi_p$.
We evaluate the matrix element in Eq.~(\ref{Lehmann}) in the first
order in the intrawire interaction.  The nonvanishing at $\omega-v_F k
\gg \delta\omega$ contribution results from the processes in which the
unperturbed final state $|n \rangle$ in Eq.~(\ref{Lehmann})
has two electron-hole pairs: $\left|n\right\rangle^{(0)} =
\psi_{p+q}^\dagger \psi_p \psi_{p'-q'}^\dagger
\psi_{p'}\left|0\right\rangle$.  This contribution is
\begin{eqnarray}
&& \delta S(k,\omega ) =
\frac{1}{\pi ^2} \int dpdp'dqdq' \, \delta(q-q'-k) 
\label{F}\\
&& \quad\quad
\times \;\delta (\omega -\xi _{p+q}+\xi_p-\xi_{p'-q'}+\xi_{p'}) 
\nonumber\\ 
&& \quad\quad
\times \;f_p(1-f_{p+q}) f_{p'}(1-f_{p'-q'}) K^2(p,p',q,q',\omega),
\nonumber 
\end{eqnarray}
where $f_p$ are the Fermi functions, $\xi_p = p^2/2m$, and 
\begin{eqnarray*}
K&=& \frac{U(q')-U(p-p'+q')}{\omega-\xi_{p+q}+\xi_{p+q'}} 
\;-\frac{U(q') - U(p-p'+q)}{\omega-\xi_{p+q-q'}+\xi _{p}}
\\
&&\quad +[p\leftrightarrow p' , q \leftrightarrow - q'].
\end{eqnarray*}
Note that Eq.~(\ref{F}), unlike Eq.~(\ref{Lehmann}), accounts for a
finite temperature.  At $\omega\ll \epsilon_F$ and $k\ll k_F$,
Eq.~(\ref{F}) yields the interaction-induced correction to
$A(k,\omega)$,
\begin{equation}
\delta A(k,\omega) 
= \frac{{\widetilde U}^2}{v_F} \frac{k^4}{m^2}
\frac{\theta (\omega - v_F k)}{\omega^2 - v_F^2 k^2},
\label{AA}
\end{equation}
where ${\widetilde U} = [U(0)-U(2k_F)]/2\pi v_F \ll 1$. This result is
valid for $\omega\ll \epsilon_F$, $k\ll k_F$, 
$|\omega-v_F k|\gg\max\{{\widetilde U} v_Fk, \delta\omega(k,T)\}$, 
and describes $A(k,\omega)$ outside the interval Eq.~(\ref{delta_free}).  
The limit of linear electron dispersion relation ($m\to\infty$)
corresponds~\cite{DL} to $\delta A(k,\omega)=0$.

We use Eq.~(\ref{AA})  to evaluate the interaction-induced
correction $\delta r$ to the drag resisitivity between non-identical wires 
with $T_1= k_F\delta v\gg \epsilon_F\widetilde{U}$. At the lowest
temperatures, Eqs.~(\ref{drag_1}) and (\ref{alpha}) yield
\begin{equation}
\delta r \sim \frac{\widetilde{U}^2}{l_0}
\left(\frac{T_1}{\epsilon_F}\right)^4 \left(\frac{T}{T_1}\right)^5.
\label{off_peak}
\end{equation}
With increasing temperature, the $r(T)$ dependence changes from
Eq.~(\ref{off_peak}) to the activation law (\ref{off_peak_free}).
At $T\gg T_1$ the difference between wires does not affect $r(T)$.

We will argue now that intrawire interactions do not change the
quadratic temperature dependence of $r(T)$ at $T_1\ll T\ll T_0$, see
Eq.~(\ref{drag_peak_free}). At these temperatures, an estimate
equivalent to Eq.~(\ref{drag_peak_free}) reads $r\sim
|U_{12}^2(0)|v_F^{-3}\delta\omega (k_T,T)$ and yields $r\propto T^2$;
here $k_T\sim T/v_F\ll k_F$ is the wave vector of a typical
electron-hole excitation. Interaction apparently does not affect the
functional dependence of $\delta\omega$ on $k$ and $T$; the estimate
(\ref{delta_free}) still can be used, although the coefficients $1/m$
and $1/k_F$ in it are affected by the interaction.

The Tomonaga-Luttinger model is insufficient for the evaluation of
$\delta\omega$ in the presence of interaction: it implies linear
electron spectrum, which yields~\cite{DL} $\delta\omega = 0$.
Accounting for the curvature of the electron spectrum complicates the
treatment of the interaction greatly. The width $\delta\omega$ can be 
explicitly evaluated in the Calogero-Sutherland model which is 
characterized by a very specific interaction potential,
\begin{equation}
U_{ii}(x) = \frac{2\pi^2}{mL^2}  
\frac{\lambda (\lambda-1)}{\sin^2\left[\pi x/L\right]}.
\label{Calogero}
\end{equation}
The parameter $\lambda$ here is related to the conventional interaction
parameter $g$ of the Luttinger liquid: $g=1/\lambda$. This relation
follows from the definition $g=v_F/u$ in terms of the velocity of the
collective mode (plasmon) $u$, and its value $u = (\pi n/m)\lambda$ 
in the Calogero-Sutherland model~\cite{SLA,Ha}. For the rational values 
of $\lambda$ and at $T=0$ the density-density correlation function is 
known exactly~\cite{SLA,Ha}. Due to the integrability of the model, 
$A_i(k,\omega)\neq 0$ only in a finite interval of $\omega$ around 
$\omega = u_ik$~\cite{tail}. We found this interval for $k \leq 2\pi n_i$:
\begin{equation}
- (1/g) \frac{k^2}{2m} < \omega - u_i k < \frac{k^2}{2m}, 
\label{borders}
\end{equation}
which yields for the width
\begin{equation}
\delta\omega(k,0) = \frac{1+g}{2g} \frac{k^2}{m}.
\label{interval}
\end{equation}

In order to estimate $r$ we note that 
Eq.~(\ref{drag}) and the sum rule (\ref{sum_rule}) remain valid 
in the presence of interactions within the wires. This allows us 
to follow the steps that led to Eq.~(\ref{drag_peak_free}). Replacing 
$v_F$ by the plasma velocity $u$ in Eq.~(\ref{drag_1}) and using 
Eq.~(\ref{interval}), we find  
\begin{equation}
r = \frac{c_g}{ l_0} \left(\frac{T}{\epsilon_F}\right)^2,
\quad
c_g \propto
\frac{g^6}{1+g},
\label{CSM}
\end{equation}  
which agrees with our expectation for the $r(T)$ dependence. We are
not aware of a reliabale theory of $A_i(k,\omega)$ beyond the exactly
solvable case. However, the self-consistent Born approximation
results~\cite{Samokhin} allow us to corroborate the estimate
$\delta\omega\propto k^2/m$ for the peak width, so, apparently, the
$r\propto T^2$ dependence is universal.

%%%%%%%%%%%%%%%%%%%%%%%%%%%%%%%%
\begin{figure}[h]
\includegraphics[width=0.75\columnwidth]{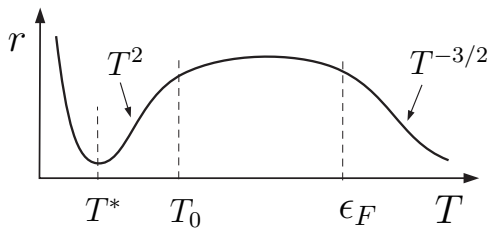}
\caption{Sketch of the temperature dependence of the drag resistivity 
between identical wires.  The small momentum transfer contribution 
considered in this paper dominates at $T>T^*$; the ratio 
$T^*/\epsilon_F$ is exponentially small for $k_F d >1$ .
\label{overall}
}
\vspace{-2mm}
\end{figure}
%%%%%%%%%%%%%%%%%%%%%%%%%%%%%%%%

First observations of drag between ballistic quantum wires appeared 
recently~\cite{exp1,exp2}. In a limited temperature interval, 
$0.2~{\rm K}< T <0.9~{\rm K}$, a three-fold drop in the drag resistance 
was observed~\cite{exp1} with the increase of temperature.
This drop was fit to a power-law $r\propto T^{-0.77}$ and interpeted
as evidence of the Luttinger liquid behavior. However, the Fermi wave
vector in the wires of Ref.~\cite{exp1} was estimated to be $k_F =
6\times 10^{4}~{\rm cm}^{-1}$, which yields $\epsilon_F=\hbar^2
k_F^2/2m^*\approx 0.2~{\rm K}$ (we used here $m^* = 0.068\,m_0$ 
known for GaAs). It thus appears that the measurements of Ref.~\cite{exp1}
correspond to a non-degenerate or weakly degenerate regime
incompatible with the Luttinger liquid description. An alternative
explanation of the observations~\cite{exp1,exp2} is provided by our
theory. Indeed, using the values of $k_F$ and $d = 200~{\rm nm}$
of~\cite{exp1}, we find $k_F d = 1.2$. Under this condition the small
momentum transfer contribution dominates at $T > T^*$, see
Fig.~\ref{overall}. The observed~\cite{exp1,exp2} behavior of $r(T)$
may correspond to the crossover regime between the limits $r(T)= \rm
const$ and $r(T)\propto T^{-3/2}$ presented by Eqs.~(\ref{high_T})
and~(\ref{hight}).

To conclude, the small momentum transfer contribution dominates 
Coulomb drag at almost all temperatures if the distance between the 
wires exceeds the Fermi wavelength, see Fig.~\ref{overall}.
Drag by small momentum transfer is possible because electron
dispersion relation is not linear, and therefore can not be accounted
for in the conventional Tomonaga-Luttinger model.

\begin{acknowledgments}
We thank KITP, Santa Barbara, for hospitality, and K. Flensberg, 
A. Kamenev, S. Tarucha, and A.  Tsvelik for discussions. This work was
supported by NSF grants DMR97-31756, DMR-9984002, DMR02-37296,
EIA02-10736, and by the Packard Foundation.
\end{acknowledgments}


\begin{thebibliography}{0}
\bibitem{tail} 
For a generic interaction $A_i(k,\omega)$ has a tail $\delta A_i(k,\omega)$ 
extending outside the interval (\ref{borders}), see Eq.~(\ref{AA}). 
The corresponding contribution to the integral in Eq.~(\ref{sum_rule}) 
scales as $k^4$ at small $k$. Accordingly, the presence of the tail does 
not affect the estimate of the height of the peak in $A_i(k,\omega)$, and, therefore, 
the $r\propto T^2$ 
dependence.
\end{thebibliography}
\end{document}